\def\be{\begin{equation}}
\def\ee{\end{equation}}
\def\bea{\begin{eqnarray}}
\def\eea{\end{eqnarray}}

\documentclass{ws-mpla}

\begin{document}

\markboth{Taotao Qiu} {Theoretical Aspects of Quintom Models}

\catchline{}{}{}{}{}

\title{Theoretical Aspects of Quintom Models}

\author{\footnotesize Taotao Qiu}

\address{Physics Department, Chung-Yuan Christian University, Chung-li,
Taiwan 320\\
qiutt@ihep.ac.cn}

\maketitle

\pub{Received (Day Month Year)}{Revised (Day Month Year)}

\begin{abstract}
Quintom models, with its Equation of State being able to cross the
cosmological constant boundary $w=-1$, turns out to be attractive
for phenomenological study. It can not only be applicable for dark
energy model for current universe, but also lead to a bounce
scenario in the early universe.

\keywords{Keyword1; keyword2; keyword3.}
\end{abstract}

\ccode{PACS Nos.: include PACS Nos.}

\section{introduction}
For decades of years, the observational data has put strong
evidences on existence of Dark Energy. The earliest evidence comes
from the observation of Type Ia Supernovae by {\it Supernova Search
Team (SST)} and {\it Supernova Cosmology Project (SCP)} in 1998
which discovered that our universe has been accelerating
\cite{Riess:1998cb,Perlmutter:1998np}. This accelerating requires a
kind of negative pressure matter in order to validate the Einstein
Gravity. Other observations implies that our universe is nearly flat
(with $\Omega_{total}=0.9996\pm0.0199$ where $\Omega_{total}$
denotes for the total relative energy density of our universe),
while the baryon matter and cold dark matter only takes small part
of nearly $27\%$, leaving large occupation for dark energy
\cite{Komatsu:2008hk}. Due to this reason, it is of great importance
to study the properties of dark energy. However, at the very
beginning, people will always ask the question: {\it what} is the
dark energy?

In the literature, plenty of dark energy candidates has been
proposed, see [\refcite{Copeland:2006wr}] for a review. People often
classify these candidates with respect to its Equation of State
(EoS) $w=\frac{p}{\rho}$, where $p$ and $\rho$ denotes for the
pressure and energy density, respectively. The simplest dark energy
candidate is the cosmological constant with energy density being
near the vacuum energy $\rho_{\Lambda}\approx(10^{-3}eV)^4$ without
varying with time \cite{Weinberg:1988cp}. This candidate, proposed
initially by A.Einstein, however suffers from the severe problem of
fine-tuning and coincidence. For this sake, dynamical dark energy
models were proposed, among which are Quintessence ($w>-1$)
\cite{Ratra:1987rm}, Phantom ($w<-1$) \cite{Caldwell:1999ew},
K-essence ($w>-1$ or $w<-1$)
\cite{Chiba:1999ka,ArmendarizPicon:2000dh}, Quintom ($w$ crosses
$-1$) \cite{Feng:2004ad}, etc. Moreover, it is widely realized that
accelerating can also be obtained by modifying the Einstein's
Gravity \cite{Capozziello:2003tk}. However, the new released data of
Supernovae, Wilkinson Microwave Anisotropic Probe observations
(WMAP) and Sloan Digital Structure Survey (SDSS) as well as the
forthcoming Planck etc., implies that although the cosmological
constant with $w=-1$ fits the data well, dynamical candidates still
cannot be ruled out. Specifically, the Quintom model whose EoS can
cross $-1$ is mildly favored.

Since the observational data mildly favors Quintom, people may ask:
How can we construct such a model theoretically? As is well known,
not all of the dark energy models can make it EoS cross $-1$ which
is constrained by the so called ``No-Go Theorem"
\cite{Vikman:2004dc,Hu:2004kh,Caldwell:2005ai,Zhao:2005vj,Kunz:2006wc,Xia:2007km}.
In this theorem, it is demonstrated that for theory of dark energy,
the EoS of dark enegy model will by no means cross the cosmological
constant boundary if it is (1) in $4D$ classical Einstein Gravity,
(2) described by single simple component (either perfect fluid or
single scalar field with lagrangian as ${\cal L}={\cal
L(\phi,\partial_\mu\phi\partial^\mu\phi)}$), and (3) coupled
minimally to Gravity. Indeed, the crossing for such a dark energy
model will lead to the divergence of physical quantities, which
consequently result in the inconsistency of the system. For sake of
this theorem, people proposed various kinds of Quintom models, such
as double field Quintom, non-scalar (spinor, vector, etc.) Quintom,
Quintom with higher derivative operator, non-minimally coupled
Quintom, etc. These models corresponds to violation of different
conditions mentioned in the no-go theorem. Some of those models will
be reviewed in detail in the next section. Due to the behavior of
EoS of Quintom models, it can as well lead to various evolution of
the universe which cannot be realized by non-Quintom dark energy
models.

Furthermore, if we apply the property of Quintom to the early
universe, some interesting features will also be expected, for
example, a bounce scenario will be obtained \cite{Cai:2007qw}. A
Quintom model will give rise to a bounce scenario in $4D$ Einstein
Gravity, i.e. there is no need to introduce extra dimensions. This
can be seen by investigating the conditions for a bounce to happen.
In a contracting phase the scale factor is decreasing, i.e.,
$\dot{a(t)}<0$, while in an expanding phase the scale factor is
increasing, $\dot{a(t)}>0$. Therefore we expect that at the
transition point $\dot{a(t)}=0$ while $\ddot{a(t)}>0$. Equivalently,
we expect the Hubble parameter $H$ cross the zero point from $H<0$
to $H>0$ at the bouncing point, which requires the EoS of the
universe less than $-1$ according to the Einstein Equation. After
the bounce, in order for the universe to enter into the realistic
one with matter dominating, radiation dominating, etc., it is
required that the EoS of the universe being larger than $-1$. That
is, the EoS will cross $-1$ during the whole bounce process, and a
Quintom behavior is needed.

Based on the scenario of Quintom bounce, an important problem is
that how can it give rise to the observed amount of perturbation to
form the large scale structures? To answer this question, we
constructed the perturbation theory of Quintom bounce
\cite{Cai:2007zv,Cai:2008qb}. In our model, Quintom field can act as
inflaton after bounce and drive enough period of inflation. In this
scenario, the inflation stage can dilute everything away and set the
universe to be near the state of the universe after the standard
inflation, so the bounce process may hardly effect the following
evolution of the universe and fits well with all the current data.
Nevertheless, we do expect that bounce can leave some signature that
can be seen for future observations. Furthurmore, using Quintom
field we can also build another bounce model known as ``Lee-Wick
type" or ``matter bounce", which requires no inflationary stage at
all \cite{Cai:2008qw}. In this scenario, the perturbation is also
calculated and a scale invariant power spectrum is produced during
the contraction of the universe.
\section{the Quintom model}
\begin{figure}[htbp]
\includegraphics[scale=0.5]{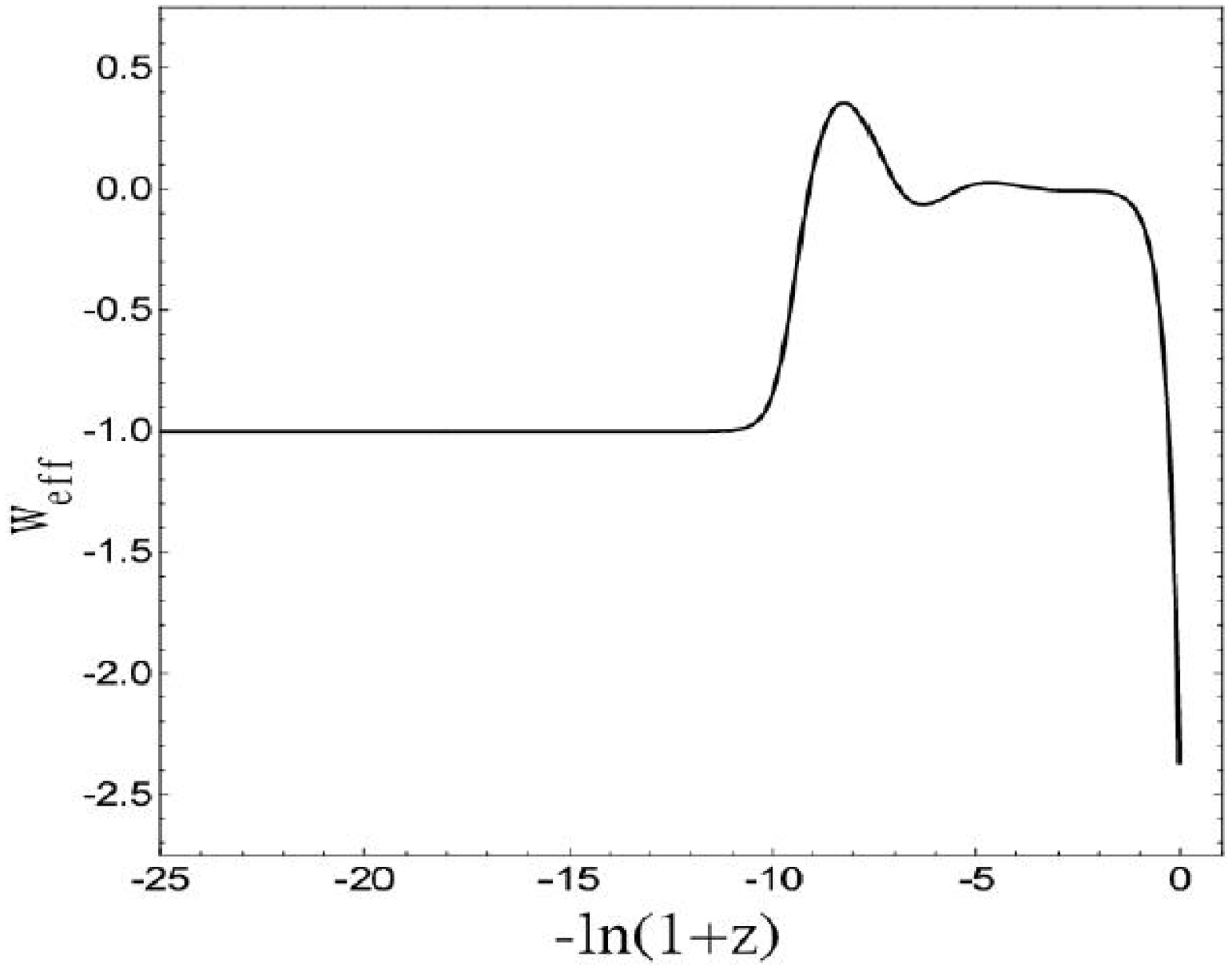}
\caption{The evolution of the effective EoS of double-field Quintom
model proposed in [10]. The parameters are chosen as:
$V_0=8.38\times10^{-126}m_{pl}^4$, $\lambda=20$. The initial
conditions were chosen as: $\phi_{1i}=-1.7m_{pl}$,
$\phi_{2i}=-0.2292m_{pl}$.} \label{quintom}
\end{figure}

Quintom model was initially proposed in [\refcite{Feng:2004ad}],
where the authors combined the Quintessence and Phantom components
together. The typical action of these kind of model is: \be {\cal
}S_{Quintom}=\int
d^4x\sqrt{-g}\{\frac{1}{2}\nabla_\mu\phi_1\nabla^\mu\phi_1-\frac{1}{2}\nabla_\mu\phi_2\nabla^\mu\phi_2-V(\phi_1,\phi_2)\}~,\ee
where $\phi_1$ and $\phi_2$ are Quintessence and Phantom components,
respectively. Due to the combined effects of the two, it is an
intuition to see that the total EoS of the whole system will evolve
across the cosmological constant boundary. Generally, the effective
potential can be of arbitrary form, while the two components can be
either coupled or decoupled. In the original paper
\cite{Feng:2004ad}, the authors considered a simple form of
decoupled case:
$V(\phi_1,\phi_2)=V_0(e^{-\frac{\lambda\phi_1}{m_{pl}}}+e^{-\frac{\lambda\phi_2}{m_pl}})$
where $m_{pl}$ denotes for the Planck mass while $\lambda$ is
dimensionless constant. The evolution of the EoS was plotted in Fig.
\ref{quintom}.

After the first Quintom paper came up, lots of people investigated
its property due to its importance. In [\refcite{Guo:2004fq}], the
attractor solution has been studied and in [\refcite{Zhang:2005eg}],
people extended to the more general case of which a coupling term
has been introduced. See also [\refcite{Cai:2009zp}] for more
variety of double-scalar Quintom models. In [\refcite{Zhao:2005vj}],
the perturbation of Quintom model has been calculated and a
self-consistent perturbation theory were constructed.

Double field is the simplest and most natural scenario of Quintom
model. However, it suffers from many problems such as big-rip and
quantum instability. So people have to think about alternative ways
to realize the EoS crossing. Another Quintom model is the addition
of a higher derivative operator to the single scalar field
\cite{Li:2005fm}. The most general lagrangian is as follows: \be
{\cal L}={\cal
L}(\phi,\nabla_\mu\phi\nabla^\mu\phi,\Box\phi,\nabla_\mu\nabla_\nu\phi\nabla^\mu\nabla^\nu\phi,...)~,\ee
where $\Box=\nabla_\mu\nabla^\mu$ is the d'Alembertian operator and
the ellipse denotes for other higher dimensional operators. The
higher order operators can be derived from fundamental theories such
as string theory or quantum gravity
\cite{Simon:1990ic,Eliezer:1989cr,Erler:2004hv}, and with the
addition of high order terms to the Einstein Gravity, the theory is
shown to be renormalizable \cite{Stelle:1976gc}. Because of the
extra degrees of freedom provided by the higher order term, it can
simulate double-field model in some specific cases. However, it has
more interesting features of its own. In Higher derivative theory,
the dispersion relation is modified, and it may provide possible
solutions to the problem of quantum instability
\cite{Simon:1990ic,Hawking:2001yt}. Furthermore, because it is more
complex and involved in the high derivative term, it may give rise
to some new behaviors of evolution of the universe.

\begin{figure}[htbp]
\includegraphics[scale=0.5]{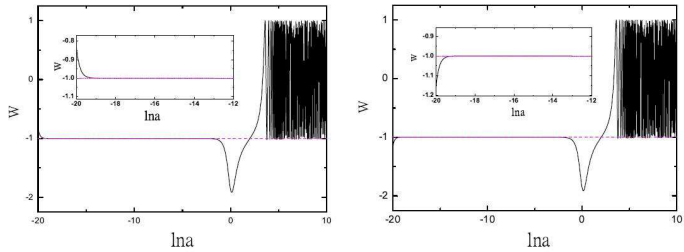}
\caption{The evolution of EoS of the model proposed in [31]. For the
left hand side case the initial values are $\psi_i=-0.26m_{pl}$,
$\dot\psi_i=3.52\times10^{-62}m_{pl}^2$, $\chi_i=0.25m_{pl}$,
$\dot\chi_i=-3.62\times10^{-62}m_{pl}^2$, while for the right hand
side case they are $\psi_i=-0.26m_{pl}$,
$\dot\psi_i=-2.84\times10^{-62}m_{pl}^2$, $\chi_i=0.25m_{pl}$,
$\dot\chi_i=2.74\times10^{-62}m_{pl}^2$.}\label{hd1}
\end{figure}

We consider the phenomenological form of higher derivative action as
follows \cite{Zhang:2006ck}: \be {\cal S}_{HD}=\int
d^4x\sqrt{-g}\{A(\phi)\nabla_\mu\phi\nabla^\mu\phi-\frac{C(\phi)(\Box\phi)^2}{m_{pl}^2}-V(\phi)\}~,\ee
where $A(\phi)$ and $C(\phi)$ are some functions of the field
$\phi$. For various choice of the form of these functions, the
evolution of the field could be very different. For sake of
simplicity, we choose $A(\phi)=-\frac{1}{2}$ and $V(\phi)=0$ as an
example. By redefining the field variable
$\chi=\frac{C(\phi)}{m_{pl}^2}\Box\phi$ and $\psi=-(\phi+\chi)$, we
obtain a simple form of action: \be {\cal S}_{eff}=\int
d^4x\sqrt{-g}\{-\frac{1}{2}\nabla_\mu\psi\nabla^\mu\psi+\frac{1}{2}\nabla_\mu\chi\nabla^\mu\chi-\frac{m_{pl}^2\chi^2}{C(\psi+\chi)}\}~,\ee
and the last term could be viewed as an effective mass term of
$\chi$ which varies in terms of $\psi$ and $\chi$. Here we can see,
$\chi$ acts as a normal field while $\psi$ being the ``ghost" field
with a wrong sign. We can choose the form of $C(\phi+\chi)$ to
control the evolution of the two field, so that the EoS of the whole
system can not only cross $-1$, but also present novel behaviors. In
our numerical analysis, we choose $C(\phi+\chi)$ to be small at the
beginning and large at the end of the evolution, both of the two
region being nearly a constant, while in the mediate region it has a
significantly running. Therefore, at the beginning of evolution the
``ghost" field $\psi$ evolves as a massless field with its own
effective EoS nearly unity. In the mediate region, however, the
effect of $\psi$ in potential term is involked and it behaves like a
real Phantom and draw the total EoS below $-1$. In the future, it
behaves like massless field again. While in this region the normal
field $\chi$ gets a large value of effective mass, the whole system
will evolve as a non-relativistic matter with the EoS oscillating
around zero. See numerical results in Fig. \ref{hd1}.

For actions of field theory linear with the kinetic term, higher
derivative operator could only exist as higher order term with an
energy cut-off. However, it can also exist with the same order as
the kinetic term in a non-linear field theory. An explicit example
is the so-called ``String Inspired Quintom", where such a term
resides in the square root in the Dirac-Born-Infeld (DBI) action
\cite{Cai:2007gs}. The effective action is given as: \be {\cal
S}_{DBI}=\int
d^4x\sqrt{-g}\{-V(\phi)\sqrt{1-\alpha'\nabla_\mu\phi\nabla^\mu\phi+\beta'\phi\Box\phi}\}~,\ee
where $\alpha'$ and $\beta'$ are coefficients of dimension 4. One
can see that without the high derivative term, the action will be
reduced to the normal DBI action describing a tachyon state in
string theory \cite{Sen:2002nu}. However, one cannot remove this
term by adding a total integral to the action as in the linear
theory. Furthermore, one can also include infinite numbers of higher
order derivative terms as in the context of p-adic string theory
\cite{Barnaby:2006hi}.

\begin{figure}[htbp]
\includegraphics[scale=0.5]{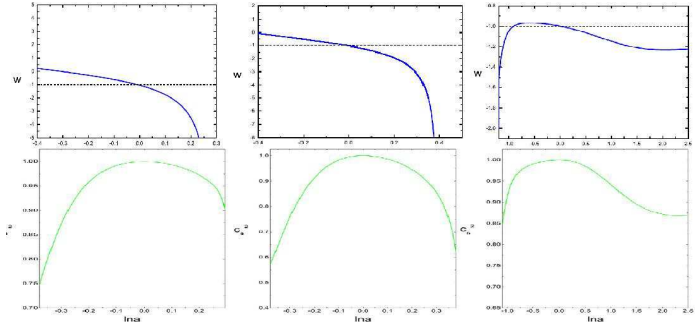}
\caption{The first row: The evolution of the EoS of the ``String
Inspired Quintom model", of which the first picture for potential
$V(\phi)=V_0e^{-\lambda\phi^2}$ while the last two for potential
$V(\phi)=\frac{V_0}{e^{\lambda\phi}+e^{-\lambda\phi}}$ and different
parameter choices. The second row: The variation of the
corresponding $c_s^2$ of each case obove.}\label{hd2}
\end{figure}

Due to the effect of the higher derivative operator, one can see
from the numerical results in Fig. \ref{hd2} that the EoS can also
cross $-1$ naturally and various behaviors are presented according
to the form of the potential. As a non-linear theory, it is also
important to check the stability of classical perturbation. We
calculated the second order action of this model and numerically
obtained the variation of $c_s^2$ with respect to time during
evolution. From the result one could see that for our cases $c_s^2$
varies between the range of $(0,1)$, denoting neither instability
nor unphysical propagation of the fluctuation.

Of course there are other models that can make EoS crossing $-1$,
among which are vector fields \cite{ArmendarizPicon:2004pm}, spinor
fields \cite{Cai:2008gk}, non-minimal coupling fields
\cite{Perivolaropoulos:2005yv}, as well as the theories of modified
gravity or high dimensional theories
\cite{Aref'eva:2005fu,Zhang:2006at}, which will not be discussed
here because of the page limit.

As a side remark, it is noticeable that due to the dynamic behavior
of Quintom models, it can lead to many interesting fates of the
universe in the future, which is expected to be determined by
forthcoming observations. The asymptotic behavior of Quintom can
mimic that of Quintessence, Phantom, as well as $\Lambda$CDM model.
Furthermore, it can bring novel features that cannot be realized by
any of them, for instance, the oscillating behavior around the
cosmological constant boundary. Within this phenomenon, we could
construct a cyclic or recurrent universe
\cite{Feng:2004ff,Xiong:2008ic}. Other features include that the
Quintom models has a cosmic self-duality where one kind of Quintom
model in expanding universe is dual to the other in contracting
universe depending on the initial conditions \cite{Cai:2006dm}. In
this sense, Quintom model is of very much interest in phenomenology.
Meanwhile, there are also many subtle issues about Quintom that
remains unclear, such as its connection to the fundemental theories
or particle physics, and its nature in quantum levels, etc., which
are worthwhile of investigation in the future.

\section{Quintom bounce story: background and perturbation}
For decades of years, the theory of inflation has attracted many
attentions for its success in solving most problems (flatness
problem, horizon problem, etc) that arised in Standard Big-Bang
Cosmology \cite{Guth:1980zm}. However, inflation is far from a
complete theory since it suffers from other problems such as
singularity problem \cite{Borde:1993xh} and transplanckian problem
\cite{Martin:2000xs}. Due to this reason, people proposed several
alternative solutions of the early universe, among which are
Pre-Big-Bang scenario \cite{Gasperini:1992em}, Ekpyrotic scenario
\cite{Khoury:2001wf}, string gas scenario
\cite{Brandenberger:1988aj}, non-local string field theory scenario
\cite{Aref'eva:2007uk} and so on. When restricted to 4D effective
theory, a simple way to get rid of the singularity is to have a
bounce process at the early stage. It can be realized by a model of
ghost condensate \cite{Creminelli:2007aq,Buchbinder:2007ad} or the
modification of Einstein Gravity
\cite{Brustein:1997cv,Cartier:1999vk,Tsujikawa:2002qc,Biswas:2005qr,Biswas:2006bs,Cai:2009in}.

\begin{figure}[htbp]
\includegraphics[scale=0.5]{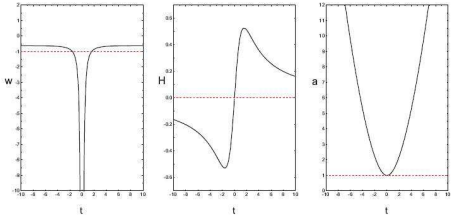}
\caption{The evolution of the EoS $w$, hubble parameter $H$ and the
scale factor $a$ as a function of the cosmic time $t$. The EoS is
parameterized as $w=-r-\frac{s}{t^2}$ where $r=0.6$ and
$s=1$.}\label{parametric}
\end{figure}

Quintom model, as mentioned in the introduction, can also provide a
bounce solution of the early universe, avoiding the singularity
naturally. As an example, In Fig. \ref{parametric} we draw a
phenomenological parametrization of Quintom model. From the picture
we can see that as the EoS evolves, the scale factor of the universe
can transfer from damping to growing, i.e., a bounce happens.

In order to compare the predictions of Quintom bounce scenario to
the observations, it is necessary to investigate the perturbations
of the scenario. Indeed, to make the scenario realistic, one need
the perturbations after the bounce as seeds of forming our galaxies
and large scale structures. In singular bounce scenarios such as Pre
Big Bang and Ekpyrosis, the fluctuation cannot evolve through the
bounce point consistently. It will diverge at the pivot and thus
invalidate the linear perturbation theory. While in our scenario, as
we will see, the proper matching condition can be used and the
fluctuations can transfer from contraction to expansion naturally.
Furthermore, according to different mode, the fluctuation can
possess features both found in singular and non-singular bouncing
models. The power spectrum of our scenario has also been calculated,
and due to different models, we can get either running or scale
invariant power spectrum \cite{Cai:2007zv,Cai:2008qb}.

In what follows, we will focus on bounce scenario caused by
non-interacting double field Quintom. The general form of such a
model is: \be {\cal
L}_{QB}=\frac{1}{2}\nabla_\mu\phi\nabla^\mu\phi-\frac{1}{2}\nabla_\mu\psi\nabla^\mu\psi-V(\phi)-W(\psi)~,\ee
where $V(\phi)$ and $W(\psi)$ are potentials for normal and ghost
fields. We will see that due to different forms of potentials,
different results will come about.

{\it Case I. $V(\phi)=\frac{1}{2}m^2\phi^2$, $W(\psi)=0$.}

In this case \cite{Cai:2007zv}, the ghost field only remains its
kinetic term, thus its energy density evolves proportional to
$a^{-6}$ and become important only near the bouncing point. At
regions far away from the bounce, the universe is dominated by the
normal field. We begin the contracting phase with $\phi$ oscillating
around the minimum of the potential. Because of the contraction of
the universe, the amplitude of the oscillation grows as $a^{3/2}$
and the field behaves as non-realistic matter. This period is called
``Heating phase". After the last oscillation, the field climb up
along the potential and caused a period of ``deflation" of which the
EoS of the universe approximately near $-1$. Meanwhile, the energy
density of the field $\psi$ is growing all the time. When it catches
up with that of the field $\phi$, the total energy density vanishes.
As we learn from Einstein Equation that $H=0$ with a positive time
derivative $\dot H$. Thus the bounce happens. After the bounce the
$\psi$ field damps quickly while the $\phi$ field rolls down slowly
along its potential, very much like the chaotic inflation model. At
last, $\phi$ oscillates around the minimum of the potential again,
with a damping amplitude.

\begin{figure}[htbp]
\includegraphics[scale=0.5]{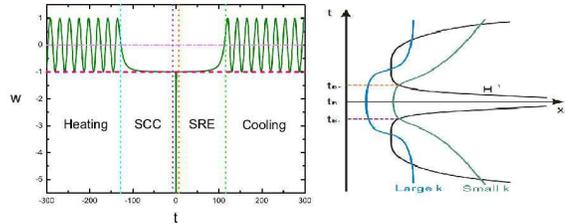}
\caption{The left hand side: The evolution of the EoS in Quintom
bounce model of {\it Case I}. The initial values are chosen as
$\phi_i=-5.6\times10^{-3}m_{pl}$,
$\dot\phi_i=2.56\times10^{-10}m_{pl}^2$,
$\dot\psi_i=4.62\times10^{-85}m_{pl}^2$ with
$m=1.414\times10^{-7}m_{pl}$. The right hand side: A sketch of the
evolution of perturbations with different comoving wave numbers $k$
in this case.}\label{bounce1}
\end{figure}

The left hand side of Fig. \ref{bounce1} is the evolution of EoS
during the bounce process and the right hand side is the sketch plot
of the space-time in bounce scenario while the horizonal and
vertical axis denotes for physical distance and time respectively.
The black solid line represents the Hubble horizon and green and
blue line are different fluctuation modes.

From perturbed Einstein Equation, we get the equation for the
Newtonian gravitational potential $\Phi$ as \cite{Mukhanov:1990me}:
\be\label{perturb} \Phi''+2({\cal
H}-\frac{\phi''}{\phi'})\Phi'+2({\cal H}'-{\cal
H}\frac{\phi''}{\phi'})\Phi-\nabla^2\Phi=8\pi G(2{\cal
H}+\frac{\phi''}{\phi'})\psi\delta\psi'~,\ee where ${\cal H}$ is the
comoving Hubble constant, and prime denotes derivative with respect
to comoving time $\eta$. For the initial condition, we set it as
Bunch-Davies vacuum as usual in the far past, when
$u\equiv\frac{a\Phi}{\phi'}\sim\frac{1}{\sqrt{2k^3}}$ and
$\Phi\sim\eta^{-3}\frac{1}{\sqrt{2k^3}}$. Here we use the fact that
at the very beginning when $w\approx 0$ we have
$\frac{\phi'}{a}\sim\eta^{-3}$. At regions far from bouncing point,
the right hand side of Eq. (\ref{perturb}) can be neglected and the
equation becomes homogeneous. It has two branches of solution at
super-Hubble region, one is growing while the other is constant. At
the regions near the bounce, we get both of the branches oscillating
with the amplitude depending on the energy scale of the bouncing.
After the bounce, we again get two branches where one is constant
while the other is damping. The solutions of each stage is matched
consistently via Deruelle-Mukhanov matching conditions
\cite{Hwang:1991an,Deruelle:1995kd}.

Unfortunately, this kind of model cannot give rise to a scale
invariant power spectrum as obtained by observation. The reason is
easy to see: In the contracting phase, the perturbation will get out
of the horizon and the amplitude of the perturbation will be raised
by the growing mode. Such a behavior corresponds to the modification
of the initial condition of perturbations at the following
inflationary stage. The power spectrum will thus have a blue tilt.
For the above reason, some alternative models need to be considered.

{\it Case II:
$V(\phi)=\frac{\lambda\phi^4}{4}(\ln\frac{|\phi|}{v}-\frac{1}{4})+\frac{\lambda
v^4}{16}$, $W(\psi)=0$.}

\begin{figure}[htbp]
\includegraphics[scale=0.5]{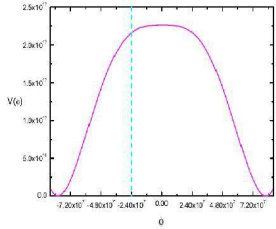}
\caption{The sketch plot of the potential as function of $\phi$ in
Quintom bounce model of {\it Case II}.}\label{potential}
\end{figure}

\begin{figure}[htbp]
\includegraphics[scale=0.5]{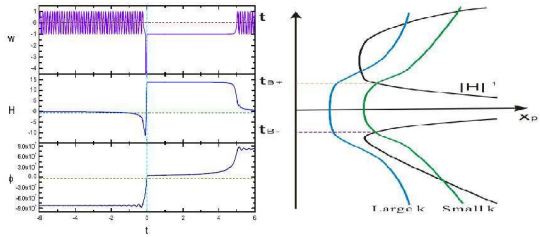}
\caption{The left hand side: The evolution of the EoS in Quintom
bounce model of {\it Case II}. The initial values are chosen as
$\phi_i=-0.82m_{pl}$, $\dot\phi_i=3.0\times10^{-10}m_{pl}^2$,
$\dot\psi_i=5.0\times10^{-13}m_{pl}^2$ with
$\lambda=8.0\times10^{-14}$ and $v=0.82m_{pl}$. The right hand side:
A sketch of the evolution of perturbations with different comoving
wave numbers $k$ in this case.}\label{bounce2}
\end{figure}

In this case \cite{Cai:2008qb}, the potential of normal field $\phi$
is of Coleman-Weinberg type, with two vacua set on both sides of the
middle \cite{Coleman:1973jx}, see Fig. \ref{potential}. We set our
initial state with $\phi$ residing in one of the vacua while
oscillating. When the amplitude grows large enough the field climbs
to the plateau of the potential, while the growing energy density of
the second field $\psi$ catches up, and the bounce happens, followed
by a period of inflation. As one could see from the numerical
analysis of background evolution that, there is no ``deflation"
period in the contracting phase and the symmetry with respect to the
bouncing point has been broken, see Fig. \ref{bounce2}. Thanks to
this symmetry violation, we could see from the sketch plot that for
large $k$ modes, the fluctuation will stay inside the horizon all
the contracting time, leading to a scale invariant power spectrum.
Meanwhile, for small $k$ modes which exits horizon during
contraction, the spectrum will have a blue tilt.

{\it Case III: $V(\phi)=\frac{1}{2}m^2\phi^2$,
$W(\psi)=-\frac{1}{2}M^2\psi^2$.}

\begin{figure}[htbp]
\includegraphics[scale=0.5]{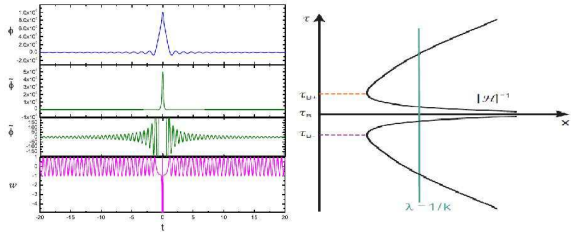}
\caption{The left hand side: The evolution of the EoS in Quintom
bounce model of {\it Case III}. The initial values are chosen as
$\phi_i=1.74\times10^{-3}m_{pl}$,
$\dot\phi_i=1.44\times10^{-8}m_{pl}^2$,
$\psi_i=8.98\times10^{-6}m_{pl}$,
$\dot\psi_i=-14.08\times10^{-12}m_{pl}^2$ with
$m=5.0\times10^{-6}m_{pl}$, $M=1.0\times10^{-5}m_{pl}$. The right
hand side: A sketch of the evolution of perturbations with different
comoving wave numbers $k$ in this case.}\label{bounce3}
\end{figure}

This is another interesting case where both of the field get a mass
term, while that of the ghost field has a wrong sign
\cite{Cai:2008qw}. This model can be obtained from the scalar sector
of Lee-Wick Standard Model with $M\gg m$ \cite{Grinstein:2007mp}. In
this case, both of the fields start off oscillating around their
extremes of the potential, and the initial stage is dominated by the
normal field. The oscillation amplitude of both fields scale as
$a(t)^{-3/2}$ and the universe presents
non-relativistic-matter-like. However, $\psi$ oscillates much
rapidly due to the heavy mass, and eventually the energy density of
$\psi$ catches up with $\phi$ and bounce happens. After the bounce,
both fields continues oscillating with a damping amplitude. During
the whole process except the bouncing point, the EoS of the universe
has an average value of $w=0$ and there is neither deflation nor
inflation era at all. See Fig. \ref{bounce3}.

In this model, a scale invariant power spectrum can be obtained
during contracting phase for fluctuations on scales larger than
Hubble radius. This could be seen easily if we work in terms of the
comoving curvature perturbation $\zeta$, or equivalently, the
Mukhanov-Sasaki variable $v=z\zeta$
\cite{Sasaki:1986hm,Mukhanov:1988jd} where $z\sim a$ for
time-independent EoS. The well-known equation of motion for $v$ is:
\be v''+(k^2-\frac{z''}{z})v=0~.\ee

For super-Hubble fluctuation modes, one can neglect the $k^2$ term
and make use of the parametrization $a(\eta)\sim \eta^2$ for $w=0$
to get $v\sim\eta^{-1}$, and hence obtain the $k$-dependence of
power spectrum: \be P_\zeta(k,\eta)\sim
k^3|v(\eta)|^2a(\eta)^{-2}\sim
k^3|v(\eta_H(k))|^2(\frac{\eta_H(k)}{\eta})^2\sim k^{3-1-2}\sim
k^0~.\ee In deriving the above formula we've used the fact that the
comoving time when fluctuations cross the Hubble horizon
$\eta_H(k)\sim k^{-1}$.
\section{summary} This talk is mainly focused on the properties
of Quintom dark energy model and its application to the early
universe. Quintom model, which needs multi degrees of freedom, has
an EoS crossing $-1$ and can bring various features of the universe
in the future. When applying to the early universe, it can give rise
to a bouncing scenario. The perturbation theory of Quintom bounce is
self consistent and a scale invariant power spectrum can be
obtained.
\section*{Acknowledgments}

It is a pleasure to thank the organizers of ``The International
Workshop on Dark Matter, Dark Energy and Matter-antimatter
Asymmetry'' for the invitation to speak and for their wonderful
hospitality in National Tsinghua University, Taiwan. I also wish to
thank Prof. R. Brandenberger, Yifu Cai, Prof. Yunsong Piao, Prof.
Hong Li, Prof. Mingzhe Li, Jie Liu, Dr. Junqing Xia, Prof. Xinmin
Zhang, Dr. Xiaofei Zhang and Dr. Gongbo Zhao for useful help. and
all the authors of the figures for allowing me to use their figures
as a citation. My research is supported in parts by the National
Science Council of R.O.C. under Grant No. NSC96-2112-M-033-004-MY3
and No. NSC97-2811-033-003 and by the National Center for
Theoretical Science.

\end{document}